# Quantifying Impacts of Wildfire Smoke on Crop Yields in California

Astrid N. Hoefler, Environmental Studies Department, University of California, Santa Cruz, California, USA
Robert B. Lund, Statistics Department Baskin School of Engineering, University of California, Santa Cruz, California, USA
J. Elliott Campbell, Environmental Studies Department, University of California, Santa Cruz, California, USA

Abbreviations: EPA, Environmental Protection Agency; PM, particulate matter; USDA, United States Department of Agriculture

## Abstract

Climate change and increasingly severe wildfires pose significant threats to both people and ecosystems, yet little research has focused on the impacts of wildfires on crops. In addition to direct effects such as health risks to agricultural workers and crop damage, there is evidence that smoke may affect crop productivity and quality. This study aims to analyze the effects of wildfire smoke on crop production in Northern California. We quantified the relationship of particulate matter on grape yields using statistical regression models with a wide range of different climate and air quality predictor variables. Our analysis identifies an association between air quality deterioration due to wildfire smoke emissions and reduction in wine grape yields. For two of the three study sites, particulate matter emerged as a significant predictor variable in combination with temperature. Final models for Sonoma and Napa had $R^2$ values of 0.79 and 0.66 respectively, suggesting a shift from climate only-controlled yields to wildfire and climate-controlled yields. This work sets the stage for the next generation of agronomic models that incorporate the growing threat of wildfire on agriculture production systems.

**Synopsis:** Little research has been done on the $PM_{2.5}$ effects on California crop yields. This study finds that wildfire driven air pollution impacts wine grape yields in Northern California.

## Introduction

Climate change is a direct threat to ecosystems, particularly with the frequency and intensity of wildfire. In high-risk fire regions, such as California, increased warming (seen statewide in California during all 12 months of the year) has exacerbated the likelihood and magnitude of hydrological drought (Diffenbaugh et al., 2015). Precipitation deficits, in combination with rising temperatures, reduced snowpack, and land-use change, not only increase the risk of extreme fire seasons, but also extend already present fire seasons (Keeley & Syphard, 2016; MacDonald et al., 2023; Mote et al., 2018; Pathak et al., 2018; Westerling & Bryant, 2008).

As the frequency and intensity of fires continue to increase, wildfire smoke is becoming a critical air quality issue. Wildfire smoke is a complex mixture comprising particulate matter (PM), carbon dioxide, water vapor, hazardous air pollutants (HAPs), and other gaseous pollutants. PM is a major component of wildfire smoke (US EPA, 2019). Wildfires are a significant source of particles with a diameter of less than 2.5 micrometers ($PM_{2.5}$) which account for approximately 90% of the total particle mass emitted from wildfires (McClure & Jaffe, 2018; US EPA, 2019). Wildfire-climate models show that wildfires have accounted for up to 50% of $PM_{2.5}$ across Western U.S. regions in recent years, compared to less than 20% a decade ago (Burke et al., 2021). On days when $PM_{2.5}$ concentrations exceed federal regulatory standards, wildfires are estimated to contribute an average of over 70% of total daily $PM_{2.5}$ (Liu et al., 2016). During the 2020 wildfires, direct $PM_{2.5}$ emissions increased average daily levels in Northern California some 38-fold (United States Environmental Protection Agency, 2025). Modeling results demonstrate that wildfires were responsible for $PM_{2.5}$ concentrations exceeding air quality standards between the months of August and October, even increasing by tenfold during the third week of August 2020 (Carreras-Sospedra et al., 2024). Future climate simulations predict significant increases in the frequency and intensity of wildfire smoke in the Western U.S. (Burke et al., 2021), suggesting a critical need to better understand wildfire smoke impacts.

While climate change and ever more destructive wildfires are widely known to impact people and ecosystems, little research has been conducted on the wildfire impacts on California's crops. In addition to direct impacts on agriculture, such as crop destruction and health impacts for agricultural workers, there is evidence that the downstream effects of wildfire smoke plumes can influence crop productivity and quality. Suspended and ground-level $PM_{2.5}$ affects the survival and physiological features of plants, which may impact plant growth and productivity (Kong et al., 2019). $PM_{2.5}$ deposited on plant leaves may block the stomata, obstructing $CO_2$ uptake for photosynthesis. Smoke also alters radiation, resulting in productivity and crop quality impacts (Hemes et al., 2020a). Increased scattered radiation and a reduction of direct radiation can also have a positive effect on plant photosynthesis (Behrer & Wang, 2022; Kanniah et al., 2012). However, once atmospheric aerosol concentrations reach a certain level, they become harmful, decreasing the photosynthetic rate of plants due to decreased radiation. Changes in photosynthetically active radiation (PAR) not only affect plant growth but also quality, including taste, appearance, and increased disease incidence (Kong et al., 2019). A regional study of the Midwestern U.S. found that high-density wildfire smoke plumes negatively affected growth for major row crops, including soybean and corn production (Behrer & Wang, 2022). In contrast, eddy flux observations in the California Central Valley, including alfalfa and wetland ecosystems, identified periods in which wildfire smoke increased diffuse radiation and enhanced ecosystem productivity (Hemes et al., 2020a). In wine grapes, wildfire smoke exposure may also affect crop quality through smoke taint. Previous studies have shown wildfire smoke exposure can lead to smoke taint in wine grapes, affecting fruit quality and marketability (Kennison et al., 2009; Sheppard et al., 2009; Summerson et al., 2021).Wildfire exposure has also been associated with visible grape leaf damage induced by ozone (Dungey et al., 2011; Kennison et al., 2009; Sheppard et al., 2009; Shertz et al., 1980). Exposure thresholds and dose duration relationships with smoke taint and yield impacts remain uncertain. However, current research suggests that even brief exposure smoke exposure during critical growing stages may induce smoke taint, indicating a rapid absorption of smoke derived compounds into the grapevine tissues (Kennison et al., 2009; Madhusoodanan, 2021; Mirabelli-Montan et al., 2021). Together these studies suggest that wildfire smoke may influence crop productivity through multiple pathways, including changes in solar radiation and air quality conditions. However, these relationships likely vary across crop types, environmental conditions, and timing of exposure.

While smoke impacts on plant growth have been demonstrated for corn, wheat, and alfalfa in the Midwestern U.S., little is known about smoke's impacts on crops in California, where wildfire frequency is rapidly increasing (California Department of Forestry and Fire Protection, 2021; Goss et al., 2020). With a $50 billion agricultural industry, air pollution in California may impede economic productivity. To date, there is a critical knowledge gap for understanding impacts of wildfires on crops in the Western U.S., including California, which produces over one third of the nation's vegetables and three quarters of its fruits and nuts.

The correlation between temperature and smoke makes disentangling the effect on crops challenging. For instance, the extreme wildfire season of 2020, which included several of the largest megafires in California history by area burned, coincided with anomalously high temperatures across California (CAL FIRE, 2024; National Centers for Environmental Information, 2020). This overlap is expected, as elevated temperatures are one of the drivers of wildfire activity. To address this complexity, statistical methods are needed to identify the contributions of temperature and smoke to crop yields.

Here, we estimate the effect of wildfire smoke on crop yields using county level regression models with meteorological and particulate matter data. We focus on wine grapes, which are one of California's leading commodities (second to dairy) and are located in regions with moderate to very high fire risk (California Department of Forestry and Fire Protection, 2023). California makes 81% of all U.S wine, making the state the world's fourth leading wine producer ($43.6 billion industry in 2021) (*CDFA - Statistics*, 2024; Davidson et al., 2025). Simulated grape yields from regression models using temperature and ambient ozone pollution as explanatory variables have successfully described crop yields (Hong et al., 2020a; Lobell et al., 2006). This is consistent with previous work that views grapes as a moderately ozone-sensitive crop (Fuhrer, 2009; Hong et al., 2020a). More broadly, regression approaches using climate variables are commonly employed in crop yield modeling (Guntukula, 2020; Hong et al., 2020b; Leng & Huang, 2017; Lobell et al., 2007a). Building on these approaches, the present study uses regression models relating wine grape yields to climatic conditions. The models additionally incorporate wildfire smoke variables to evaluate the relationship between smoke related air quality conditions and observed crop yields. Here, we use $PM_{2.5}$ as a proxy for wildfire smoke; it serves as a reliable indicator of the fine particulate matter predominantly emitted during wildfire events.

## Materials and Methods

### Study region and period

In the current study, we focus on Sonoma, Napa, and Mendocino counties in California's northern coastal region, where the leading crop is wine grapes (Figure 1). Major wine grape varieties grown across the study region include Cabernet Sauvignon, Chardonnay, Pinot Noir, and Zinfandel although varietal composition differ among counties (Whitnall & Alston, 2025; Wine Institute, 2023). The region is characterized by a Mediterranean climate conditions with warm, dry summers and cool, wet winters (Null & Mogil, 2010). While the three counties border each other, they differ in topography and exposure to marine influence. Napa County is characterized by its valley system, while Sonoma County is a more heterogenous mix of valley, coastal, and mountainous environments. Mendocino County spans a larger elevation range and includes more extensive steep forested areas than Sonoma and Napa counties. Sonoma and Mendocino experience the strongest coastal influence, due to marine fog and onshore airflow from the Pacific Ocean (Johnstone & Dawson, 2010). Napa's more enclosed valley structure limits its marine influence. Across the regions, soils are derived from volcanic, alluvial, and marine sedimentary materials (Geologic Map and Map Database of Western Sonoma,

Northernmost Marin, and Southernmost Mendocino Counties, California, 2002). Their distribution varies with local terrain and therefore also influences drainage, depth, and growing conditions. We analyze the 2010–2021 period because it provides a complete and internally consistent set of yield, $PM_{2.5}$, and climate data across all three counties. This interval also captures the full envelope of interannual variability in both climate conditions and wildfire-driven $PM_{2.5}$ extremes, including multiple major fire seasons. While newer data could support future extensions of this work, the 2010–2021 period is well suited for evaluating the relationships examined here. .

**Ground based air pollution data**

Daily maximum $PM_{2.5}$ measurements were obtained from the Environmental Protection Agency (EPA) Air Quality System (AQS) from monitoring stations for all three counties (EPA, 2021; United States Environmental Protection Agency, 2025). The monitoring sites measure $PM_{2.5}$ concentrations with an FRM sampler (R & P Model 2025 $PM_{2.5}$ Sequential Air Sampler w/VSCC - Gravimetric). The filter based sequential sampler provides 24-hour cumulative $PM_{2.5}$ mass concentration measurements in ambient conditions using gravimetric analyses (Tai, 2020). For all counties, we calculated the average monthly $PM_{2.5}$ and a monthly maximum $PM_{2.5}$ observation (the largest observation over all days in the month). We also calculated smoke days, defined as the number of days in the month where $PM_{2.5}$ measurements exceed 20 $\mu g/m^3$ (SD20) and 50 $\mu g/m^3$ (SD50), respectively. Furthermore, we calculated smoke wave episodes as defined by Liu et al. (2016) as the number of times in a month that two or more consecutive days have $PM_{2.5}$ concentrations exceeding the 20 $\mu g/m^3$ or 50 $\mu g/m^{10}$ thresholds. These smoke wave variables allow us to study more short-lived characteristics of wildfire $PM_{2.5}$ effects that cannot be captured as effectively with seasonal or monthly means. For Napa County, continuous $PM_{2.5}$ data was only consistently available from 2013 to 2020. To fit the regression model for Napa, Sonoma $PM_{2.5}$ data from 2010 to 2021 were used with Napa weather data. Napa and Sonoma are geographically adjacent and exhibit similar temporal $PM_{2.5}$ variability during overlapping observation periods, with a pooled correlation across matched daily observations of r = 0.84 (Supplemental Materials Figure 1). Counties were analyzed separately because yield and climate variables were compiled at the county level, enabling county specific model calibration and intercounty comparison.

**Crop yield data**

Annual wine grape yields at the county level between 2010 and 2021 were obtained from the California County Commissioner's Reports from the United States Department of Agriculture National Agricultural Statistics Service (USDA NASS). These data represent total county level wine grape production, aggregated across all vineyards and grape varieties within each county. The reports document harvested yields, therefore reported values may reflect environmental conditions, management decisions, and direct fire damage. Environmental conditions include heat, drought, and smoke exposure. Management decisions include abandonment of vineyard blocks due to concerns of smoke taint and degraded fruit quality. These processes are aggregated within the available data. Data for crop production, acreage, and total crop value were obtained through NASS on a county level. Vines are perennial crops with productive years of upwards of 25 years. After planting, they typically do not reach productive maturity until year five or six (Ashenfelter & Storchmann, 2016). The common grapevine (Vitis vinifera) is characterized by several growth stages. For the Northern Hemisphere, bud breaks typically begins in March and April, then bloom in May and June, followed by berry growth and coloring in July and August, and maturation in September and October (Ashenfelter & Storchmann, 2016). Harvest typically occurs anytime between August to late October and early November and depends heavily on wine grape variety. Winter dormancy occurs after autumn leaf fall (Dougherty, 2012).

**Meteorological data**

For all three study counties monthly total precipitation, minimum monthly temperature, average monthly temperature, and maximum monthly temperatures were collected from the PRISM Climate Group (PRISM Climate Group, 2024). This database has a 4x4 km grid cell resolution for each location over the 2010 to 2021 study period. The major crop growth areas in the northern coastal region of California for wine grapes are tightly clustered within Sonoma, Napa, and Mendocino counties.

### Yield regression model

To have statistically reliable models that can be extrapolated beyond the range of the data, we focused on linear regressions, representing an average rate of change. The models include meteorological data that relate crop yield to temperature as well as $PM_{2.5}$ variables:

$$Yield_t = a_0 + a_1 t + a_2 T_{aug,\,t} + a_3 T_{sep,\,t} + a_4 T_{harv,\,t} + a_5 PM_{2.5,\,t} + \varepsilon_t \qquad (1)$$

where $Yield_t$ is annual crop yield in year $t$ at site $s$, and at site $s$, $T_{aug,t}$ and $T_{sept,t}$ are the monthly maximum (largest daily high) August and September temperatures, $T_{harv,t}$ is the average maximum temperatures of August and September (over all days), and $PM_{2.5}$ is the average of the monthly maximum $PM_{2.5}$. Finally, the parameters $a_0,...,a_5$ are regression coefficients and $\varepsilon_{s,t}$ is zero mean random model error. Predictor variables were included in the regression if their p-values were below 0.10 (Table 1). To identify the most relevant predictor variables, linear regressions of the weather variables were fitted for each month and every site. Backward stepwise eliminations were conducted to eliminate insignificant factors. For the backwards regression, the significance level was set to 90%, and factors with p-values greater than 0.10 were considered for elimination at each step. We found that average maximum particulate matter and maximum temperature data were the dominant factors in such crop yield regressions. Minimum temperature, average temperature, precipitation, SD20, SD50, and smoke waves did not appreciably improve model fits and where not explored further.

## Results

### Impact of climate on the variability of crop yield

Despite relatively stable harvested acreage across the study period, there is substantial interannual variability in the wine grape yields across the three study counties (Figure 2, Supplemental Materials Figure 2).The strongest relationship between yield and temperatures occurred during August and September, which coincide with peak fire season activity. We examined additional climate and PM variables, namely minimum temperature, average temperature, precipitation, number of smoke days per month exceeding 20 $\mu g/m^3$ (SD20), number of smoke days per month exceeding 50 $\mu g/m^3$ (SD50), and smoke wave episodes. These additional factors did not appreciably improve our model fits and hence were not explored further. Using stepwise regression, we found that these variables had little effect on yield.

Figure 3 illustrates interannual and spatial variability in August and September maximum temperatures and $PM_{2.5}$ concentrations across the three study counties. Napa county consistently experienced the highest August and September temperatures during the study period. Sonoma county generally exhibited lower August and September maximum temperatures, while Mendocino county experienced relatively high August maximum temperatures but comparatively lower September maximum temperatures (particularly after 2014).

**Impact of $PM_{2.5}$ pollution on the variability of crop yields**

While climate variables are generally the focus of statistical yield studies, here we explore the potential role of air quality predictor variables: monthly average $PM_{2.5}$, monthly maximum $PM_{2.5}$, SD20, and SD50 on the yields. For Sonoma and Napa, monthly average $PM_{2.5}$ was a significant explanatory variable at level 90%. We conducted forward selection for monthly average $PM_{2.5}$, monthly maximum $PM_{2.5}$, SD20, and SD50 variables and found that monthly maximum $PM_{2.5}$, SD20, and SD50 did not appreciably improve model fit and were not included further as significant variables. Mendocino model fits concluded that all air quality predictor variables were insignificant.

To disentangle the effects of temperature and smoke on wine grape crop yields, several versions of our regression model were considered. In a temperature-only model that included the meteorological variables of minimum temperature, average temperature, and maximum temperature, we found that minimum and average temperature were not statistically significant predictands. We then performed both forward and backward regressions for a weather model that examined maximum temperature across all 12 months of the year. This analysis revealed that months outside the summer and fall seasons (specifically, May, October, and November) were insignificant predictands.

We also conducted a sensitivity analysis by fitting a model incorporating both weather and $PM_{2.5}$ concentrations in which catastrophic wildfire years (high smoke years) were removed from in each study site. Model performance declined when catastrophic fire years were excluded, suggesting that $PM_{2.5}$ concentrations may contribute to explaining interannual variability in wine grape crop yields. When performing backward regressions for temperature and $PM_{2.5}$ concentration variables during May, August, September, October, and November, May, October, and November did not have significant p-values for inclusion.

Figure 3 illustrates $PM_{2.5}$ concentrations varied across counties and years. Mendocino county experienced the highest average August $PM_{2.5}$ concentrations among the study sites in the 2018 wildfire season, coinciding with the Mendocino Complex fire. Sonoma and Napa counties experienced their highest average August $PM_{2.5}$ concentrations during the 2020 wildfire season. Results show that wine grape crop yields for Napa and Sonoma decrease with increasing average $PM_{2.5}$ (Table 2). While previous studies identify temperature as a key predictor variable for yield (Lobell et al., 2007b; Santos et al., 2011), the backwards elimination and forward selection procedures used showed that temperature and $PM_{2.5}$ variables in tandem were preferred in two of the primary wine production counties. This is not necessarily contradictory, as previous studies focused on earlier periods that had relatively low rates of wildfires. These results suggest a shift in state from yields that are controlled by temperature to a state in which yields are influenced by temperature and wildfire smoke. All models have explanatory power, with the Sonoma model including both $PM_{2.5}$ and temperature variables demonstrating the best fit with an $R^2$ of 0.79 (Table 2).

**Fire concentration model results are robust across most study areas**

Adding air quality data to the yield model substantially improved model fit when compared with models that only used temperature as a predictor variable (Table 2). The addition of the air quality data increased the $R^2$ measure of association by 27% for Sonoma and 53% for Napa.

In Sonoma, the model including both $PM_{2.5}$ and temperature variables had a large increase in $R^2$ relative to the model including only temperature variables. The Sonoma $PM_{2.5}$ with temperature model had the best fit for all study sites with an $R^2$ of 0.79 and an adjusted $R^2$ of 0.71 (Figure 2, Table 2). This model was able to follow the data's general structure as well as the extreme low yield yields that occurred during high fire activity years, particularly in the extreme year of 2020, where the temperature-only model could not.

In Napa, the model including both $PM_{2.5}$ and temperature variables was statistically superior to a temperature only model (Figure 2, Table 2). The $PM_{2.5}$ with temperature model was able to adequately predict the data as well as the extreme yields throughout the study period, particularly in recent years of extreme fire activity. For Napa and Sonoma, air quality data shows that despite the high maximum $PM_{2.5}$ in 2020, the monthly average was not the highest on record, suggesting that these sites may not have had the same level of wildfire smoke and yield impacts as Mendocino (Figure 3).

For the Mendocino site, adding air quality predictor data to the yield model did not result in substantial improvements relative to models using temperature variables alone (Figure 3, Table 2). The temperature-only model produced a statistically significant fit to the observed yields. Unlike the Sonoma and Napa sites, Mendocino's lowest yield year did occur during the extreme 2020 wildfire season. In addition, Mendocino's highest average monthly $PM_{2.5}$ concentration occurred during the 2018 wildfire season. Together, these findings suggest that the relationship between $PM_{2.5}$ concentrations and yield variability differed between Mendocino and the other sites. However, the factors contributing to these differences cannot be disentangled within the scope of this study.

## Discussion and Conclusions

This study developed a model for grape yields using climate and smoke concentration data. Our analysis identified a relationship between wildfire smoke related air quality impacts and wine grape yield variability for two of our three study sites. This relationship extends beyond variability explained by climate variables alone. For Napa and Sonoma counties, models incorporating wildfire smoke concentration data more effectively captured large interannual yield fluctuations (factor of two), particularly during catastrophic fire years. Including $PM_{2.5}$ variables also improved model performance during extreme fire years, particularly in 2020, relative to temperature only models. Overall, the results suggest a negative relationship between $PM_{2.5}$ concentrations and wine grape yields.

Although $PM_{2.5}$ originates from various sources such as urban emissions, industrial activities, and transportation, the exceptionally high levels during August and September are predominantly due to wildfires. $PM_{2.5}$ serves as a reliable indicator of wildfire smoke and provides an accurate representation of wildfire activity, particularly in California. While $PM_{2.5}$ levels across the United States have generally declined over the past few decades due to stricter environmental regulations on point sources (e.g. vehicle emissions), wildfire-prone regions have shown an opposite trend (Aguilera et al., 2021; Ford et al., 2018; McClure & Jaffe, 2018). Increases in $PM_{2.5}$ emissions, particularly in the Northwest, have been linked to wildfire activity, with the most pronounced effects occurring during the summer months (Aguilera et al., 2021; McClure & Jaffe, 2018; O'Dell et al., 2019). Studies indicate that EPA-monitored $PM_{2.5}$ concentrations were some 119% higher on some wildfire days compared to non-fire days (Liu & Peng, 2019). Additionally, substantial increases in $PM_{2.5}$ exposure from wildfire smoke are predicted, particularly in the Pacific Northwest and Northern California (Liu & Peng, 2019).

The substantial interannual variability observed in wine grape yields over the study period and across sites suggest multiple environmental and management related factors likely contributed to year to year fluctuations in production. Multiple low yield years were observed across study sites, including 2011, 2015, 2020, and 2021. This suggests that reduced yields likely reflect a combination of environmental conditions and management practices rather than wildfire activity alone. Earlier low yields, such as 2011 and 2015, occurred during differing environmental conditions. For example, 2015 coincided with major regional wildfire activity including the Valley and Rocky fires (CAL FIRE, 2026). In 2011, the region was marked by cooler growing conditions and heavy October rainfall, conditions that are favorable for increased fungal disease pressure in vineyards (Caffarra et al., 2012; NOAA, 2011). Napa and Sonoma, the two largest wine grape producing counties, experienced their lowest yields in the study period during the extreme 2020 wildfire season. The 2020 wildfire season included several of the largest fires in state history (acres burned). This included the LNU Lightning Complex, the Glass fire, and the August Complex fire which combined burned over 1.42 million acres across Sonoma, Napa, Mendocino and surrounding counties (Supplemental Materials Figure 3) (CAL FIRE, 2020, 2026). In Sonoma, yields declined roughly 35% from 4.00 tons/acre in 2019 to 2.61 tons/acre in 2020. Similarly, Napa yields declined 39% from 3.62 acres/ton to 2.20 tons/acre over the same period (Figure 2). In addition, Napa experienced its highest August maximum temperatures of the study period in 2020 (33.7 C°), highlighting the overlap between extreme fire activity and anomalously warm conditions (Figure 3). Mendocino county experienced its lowest yields in 2021, representing a 15% decline from 2020 (Figure 2). This occurred despite elevated $PM_{2.5}$ concentrations during multiple wildfire seasons years, particularly during the 2018 wildfire season. Together, these observations suggest that yield variability cannot be attributed to a single driver and likely reflects a complex interaction between wildfire smoke exposure, temperature extremes, and management practices.

Our study reveals that climate conditions and wildfire activity are associated with crop yield variability across our study sites. For Mendocino, harvest temperatures (August and September) were found to be significant predictor variables for historical crop yields. This suggests that wine grape crop productivity in Mendocino may be more strongly associated with temperature, potentially due to heat stress during key phenological phases such as fruit ripening. For Sonoma and Napa, using temperature predictor variables also provided a good model fit. However, we found that including $PM_{2.5}$ as a predictor variable in addition to the temperature predictor significantly improved both model fits ($R^2$ of 27% for Sonoma and 53% for Napa, see Table 2.) However, because the yield data represent county level harvested production, the observed relationship may reflect the combined influence of wildfire smoke exposure, direct fire damage, environmental conditions, and management responses. Previous work has documented substantial wildfire related cropland and economic losses in the Western U.S., including vineyard damage and unharvested acreage during major wildfire events (Kabeshita et al., 2023). These processes cannot be independently disentangled within the current model framework. Together, these findings suggests that while temperature plays an important role in wine grape productivity, wildfire smoke may contribute to additional stress. Wildfire smoke may alter incoming solar radiation and influence photosynthesis processes through particulate depositions on leaf surfaces, potentially compounding the physiological effects of elevated temperatures. Wildfire smoke may also contribute to carryover effects across growing seasons through impact on photosynthesis, carbon storage, and inflorescence development during critical late stage periods (Ashenfelter & Storchmann, 2016; Giese et al., 2020).These processes are not captured within the current model framework but represent an important direction for future research examining wildfire smoke and vineyard productivity. In contrast to previous studies reporting enhanced productivity associated with increased diffuse radiation under low density smokey conditions (Hemes et al., 2020b), we did not observe evidence of positive relationships between wildfire smoke exposure and wine grape yields. These differences likely reflect the context dependent nature of smoke exposure effects, which may vary across ecosystem type, crop physiology, canopy structure, and timing of exposure.

Furthermore, incidence of smoke taint may have a stronger influence on grape quality than reductions in photosynthesis, primarily altering flavor characteristics rather than through photosynthetic activity itself. Results showing that $PM_{2.5}$ was not significant predictor variable in Mendocino, despite elevated $PM_{2.5}$ concentrations during major wildfire years may reflect differences in timing of exposure, topography, wine grape production systems, and localized atmospheric conditions that may buffer or amplify the relationship between smoke exposure and observed yield outcomes.

These findings highlight the need to better understand site-specific risks when thinking of climate adaptations strategies. While maximum temperature is the strongest predictor of wine grape yields for Mendocino, crop yields for Sonoma and Napa were more strongly driven by the compounded stress from both heat and $PM_{2.5}$ exposure.

We are not aware of any other empirical studies estimating the effects of $PM_{2.5}$ on California wine grapes on a county scale. However, our findings are consistent with previous studies reporting negative effects of air pollution on perennial crop systems, including ozone related reductions in crop productivity and smoke related impacts on wine quality under wildfire conditions (Hong et al., 2020a; Kennison et al., 2009; Mirabelli-Montan et al., 2021). A study looking at weather based yield forecasts for California crops modeled California's 12 most valuable perennial crops (Lobell et al., 2007b). In that study, wine grapes were accurately modeled and demonstrated that weather changes contributed to changes to wine grape yields. A study researching the impacts of climate change and ozone on California's perennial crops accurately predicted historical yields, with wine grapes being well described (Hong et al., 2020a). Here, both weather and ozone pollution variables were used and found that grapes appear to be more sensitive to ozone than many other perennial crops. These results argue that yields for most perennial crops, including wine grapes, are negatively impacted from ambient ozone (Hong et al., 2020a). In contrast to previous work, our study is able to use $PM_{2.5}$ concentrations to assess wildfire-specific impacts on wine grape crop yields in California beyond weather-only variables models. We conclude that wildfire emissions may have a negative impact on wine grape yields in northern coastal California.

Wildfires are only one of the many factors influenced by climate change that could significantly impact future crop yield. However, a better understanding of wildfire specific effects on agricultural crops is needed. Additionally, advancing management and technology strategies that adapt to climate change may be crucial. Yield models that incorporate particulate matter data may support management strategies. Furthermore, our model could provide a useful predictive tool for evaluating future air pollution impacts on crop yields when applied to a range of future climate and wildfire scenarios. Further research is needed to investigate the impact of other components of wildfire smoke, such as HAPs and $O_3$, on crop yields.

Our study leverages statistical crop models derived from historical yields, climate data, and air quality data to contribute to a growing body of research examining the relationships between wildfire smoke exposure and agricultural production of perennial crops in California. The modeling framework also has potential applications beyond the regional scale examined here. The findings of this study may help inform future adaptation strategies, such as focusing on wine grape production systems associated with earlier harvest timing or less extractive process methods, such as grapes used for white, rosé, and sparkling wine production. The results also highlight the potential importance of month-specific mitigation strategies during periods of elevated wildfire smoke exposure. These insights may additionally help inform crop and fire insurance policies tailored to seasonal wildfire risk. Overall, this study highlights the need to better understand the magnitude of wildfire impacts on wine grapes and other agricultural production in California, an increasingly high-risk, wildfire prone region.

## References

Aguilera, R., Corringham, T., Gershunov, A., & Benmarhnia, T. (2021). Wildfire smoke impacts respiratory health more than fine particles from other sources: Observational evidence from Southern California. *Nature Communications*, *12*(1), 1493. https://doi.org/10.1038/s41467-021-21708-0

Ashenfelter, O., & Storchmann, K. (2016). Climate Change and Wine: A Review of the Economic Implications. *Journal of Wine Economics*, *11*(1), 105–138. https://doi.org/10.1017/jwe.2016.5

Behrer, A. P., & Wang, S. (2022). *Current Benefits of Wildfire Smoke for Yields in the US Midwest May Dissipate by 2050*. 1–45. https://doi.org/10.1596/1813-9450-9953

Burke, M., Driscoll, A., Heft-Neal, S., Xue, J., Burney, J., & Wara, M. (2021). The changing risk and burden of wildfire in the United States. *Proceedings of the National Academy of Sciences*, *118*(2), 1–6. https://doi.org/10.1073/pnas.2011048118

Caffarra, A., Rinaldi, M., Eccel, E., Rossi, V., & Pertot, I. (2012). Modelling the impact of climate change on the interaction between grapevine and its pests and pathogens: European grapevine moth and powdery mildew. *Agriculture, Ecosystems & Environment*, *148*, 89–101. https://doi.org/10.1016/j.agee.2011.11.017

CAL FIRE. (2020). *2020 Fire Season Incident Archive*. CAL FIRE. https://www.fire.ca.gov/incidents/2020

CAL FIRE. (2024). *Top 20 Largest California Wildfires*. https://34c031f8-c9fd-4018-8c5a-4159cdff6b0d-cdn-endpoint.azureedge.net/-/media/calfire-website/our-impact/fire-statistics/top-20-largest-ca-wildfires.pdf?rev=fba7bfc52eab4d5d87fbee5bd9416ed8&hash=270E810A7FCF091122EE2A18EB24ACB6

CAL FIRE. (2026). *CAL FIRE Incidents*. CAL FIRE. https://www.fire.ca.gov/incidents

California Department of Forestry and Fire Protection. (2021). *Top 20 Most Destructive California Wildfires*. CALFIRE. https://www.fire.ca.gov/media/t1rdhizr/top20_destruction.pdf

California Department of Forestry and Fire Protection. (2023). *Fire Hazard Severity Zones Maps | OSFM*. https://osfm.fire.ca.gov/what-we-do/community-wildfire-preparedness-and-mitigation/fire-hazard-severity-zones/fire-hazard-severity-zones-maps-2022

Carreras-Sospedra, M., Zhu, S., MacKinnon, M., Lassman, W., Mirocha, J. D., Barbato, M., & Dabdub, D. (2024). Air quality and health impacts of the 2020 wildfires in California. *Fire Ecology*, *20*(1), 6. https://doi.org/10.1186/s42408-023-00234-y

*CDFA - Statistics*. (2024, May 22). https://www.cdfa.ca.gov/Statistics/

Davidson, B. A., Black, J. A., & Cagle, W. R. (2025). Economic Disruptions, Entrepreneurial Opportunities, and the Wine Country of Northern California. *Wine Business Journal*. https://doi.org/10.26813/001c.129691

Diffenbaugh, N. S., Swain, D. L., & Touma, D. (2015). Anthropogenic warming has increased drought risk in California. *Proceedings of the National Academy of Sciences*, *112*(13), 3931–3936. https://doi.org/10.1073/pnas.1422385112

Dougherty, P. H. (2012). *The Geography of Wine: Regions, Terroir and Techniques*. Springer Science & Business Media.

Dungey, K. A., Hayasaka, Y., & Wilkinson, K. L. (2011). Quantitative analysis of glycoconjugate precursors of guaiacol in smoke-affected grapes using liquid chromatography–tandem mass spectrometry based stable isotope dilution analysis. *Food Chemistry*, *126*(2), 801–806. https://doi.org/10.1016/j.foodchem.2010.11.094

EPA. (2021). *Air Quality System (AQS) API* [Data & Tools]. https://aqs.epa.gov/aqsweb/documents/data_api.html

Ford, B., Val Martin, M., Zelasky, S. E., Fischer, E. V., Anenberg, S. C., Heald, C. L., & Pierce, J. R. (2018). Future Fire Impacts on Smoke Concentrations, Visibility, and Health in the Contiguous United States. *GeoHealth*, *2*(8), 229–247. https://doi.org/10.1029/2018GH000144

Fuhrer, J. (2009). Ozone risk for crops and pastures in present and future climates. *Naturwissenschaften*, *96*(2), 173–194. https://doi.org/10.1007/s00114-008-0468-7

*Geologic map and map database of western Sonoma, northernmost Marin, and southernmost Mendocino counties, California*. (2002). https://doi.org/10.3133/mf2402

Giese, G., Velasco-Cruz, C., & Leonardelli, M. (2020). *Grapevine phenology: Annual growth and development*.

Goss, M., Swain, D. L., Abatzoglou, J. T., Sarhadi, A., Kolden, C. A., Williams, A. P., & Diffenbaugh, N. S. (2020). Climate change is increasing the likelihood of extreme autumn wildfire conditions across California. *Environmental Research Letters*, *15*(9), 094016. https://doi.org/10.1088/1748-9326/ab83a7

Guntukula, R. (2020). Assessing the impact of climate change on Indian agriculture: Evidence from major crop yields. *Journal of Public Affairs*, *20*(1), e2040. https://doi.org/10.1002/pa.2040

Hemes, K. S., Verfaillie, J., & Baldocchi, D. D. (2020a). Wildfire-Smoke Aerosols Lead to Increased Light Use Efficiency Among Agricultural and Restored Wetland Land Uses in California's Central Valley. *Journal of Geophysical Research: Biogeosciences*, *125*(2), 1–21. https://doi.org/10.1029/2019JG005380

Hemes, K. S., Verfaillie, J., & Baldocchi, D. D. (2020b). Wildfire-Smoke Aerosols Lead to Increased Light Use Efficiency Among Agricultural and Restored Wetland Land Uses in California's Central Valley. *Journal of Geophysical Research: Biogeosciences*, *125*(2), e2019JG005380. https://doi.org/10.1029/2019JG005380

Hong, C., Mueller, N. D., Burney, J. A., Zhang, Y., AghaKouchak, A., Moore, F. C., Qin, Y., Tong, D., & Davis, S. J. (2020a). Impacts of ozone and climate change on yields of perennial crops in California. *Nature Food*, *1*(3), 166–172. https://doi.org/10.1038/s43016-020-0043-8

Hong, C., Mueller, N. D., Burney, J. A., Zhang, Y., AghaKouchak, A., Moore, F. C., Qin, Y., Tong, D., & Davis, S. J. (2020b). Impacts of ozone and climate change on yields of perennial crops in California. *Nature Food*, *1*(3), 166–172. https://doi.org/10.1038/s43016-020-0043-8

Johnstone, J. A., & Dawson, T. E. (2010). Climatic context and ecological implications of summer fog decline in the coast redwood region. *Proceedings of the National Academy of Sciences*, *107*(10), 4533–4538. https://doi.org/10.1073/pnas.0915062107

Kabeshita, L., Sloat, L. L., Fischer, E. V., Kampf, S., Magzamen, S., Schultz, C., Wilkins, M. J., Kinnebrew, E., & Mueller, N. D. (2023). Pathways framework identifies wildfire impacts on agriculture. *Nature Food*, *4*(8), 664–672. https://doi.org/10.1038/s43016-023-00803-z

Kanniah, K. D., Beringer, J., North, P., & Hutley, L. (2012). Control of atmospheric particles on diffuse radiation and terrestrial plant productivity: A review—Kasturi Devi Kanniah, Jason Beringer, Peter North, Lindsay Hutley, 2012. *Progress in Physical Geography*, *36*(2), 209–237.

Keeley, J. E., & Syphard, A. D. (2016). *Geosciences* | *Free Full-Text* | *Climate Change and Future Fire Regimes: Examples from California*. https://www.mdpi.com/2076-3263/6/3/37

Kennison, K. r., Wilkinson, K. l., Pollnitz, A. p., Williams, H. g., & Gibberd, M. r. (2009). Effect of timing and duration of grapevine exposure to smoke on the composition and sensory properties of wine. *Australian Journal of Grape and Wine Research*, *15*(3), 228–237. https://doi.org/10.1111/j.1755-0238.2009.00056.x

Kong, L., Yu, H., Chen, M., Piao, Z., Dang, J., & Sui, Y. (2019). Effects of Particle Matters on Plant: A Review. *International Journal of Experimental Botany*, *88*(4), 367–378. https://doi.org/10.32604/phyton.2019.09017

Leng, G., & Huang, M. (2017). Crop yield response to climate change varies with crop spatial distribution pattern. *Scientific Reports*, *7*(1), 1463. https://doi.org/10.1038/s41598-017-01599-2

Liu, J. C., Mickley, L. J., Sulprizio, M. P., Dominici, F., Yue, X., Ebisu, K., Anderson, G. B., Khan, R. F. A., Bravo, M. A., & Bell, M. L. (2016). Particulate air pollution from wildfires in the Western US under climate change. *Climatic Change*, *138*(3), 655–666. https://doi.org/10.1007/s10584-016-1762-6

Liu, J. C., & Peng, R. D. (2019). The impact of wildfire smoke on compositions of fine particulate matter by ecoregion in the Western US. *Journal of Exposure Science & Environmental Epidemiology*, *29*(6), 765–776. https://doi.org/10.1038/s41370-018-0064-7

Lobell, D. B., Cahill, K. N., & Field, C. B. (2007a). Historical effects of temperature and precipitation on California crop yields. *Climatic Change*, *81*(2), 187–203. https://doi.org/10.1007/s10584-006-9141-3

Lobell, D. B., Cahill, K. N., & Field, C. B. (2007b). Historical effects of temperature and precipitation on California crop yields. *Climatic Change*, *81*(2), 187–203. https://doi.org/10.1007/s10584-006-9141-3

Lobell, D. B., Field, C. B., Cahill, K. N., & Bonfils, C. (2006). Impacts of future climate change on California perennial crop yields: Model projections with climate and crop uncertainties. *Agricultural and Forest Meteorology*, *141*(2–4), 208–218. https://doi.org/10.1016/j.agrformet.2006.10.006

MacDonald, G., Wall, T., Enquist, C. A. F., LeRoy, S. R., Bradford, J. B., Breshears, D. D., Brown, T., Cayan, D., Dong, C., Falk, D. A., Fleishman, E., Gershunov, A., Hunter, M., Loehman, R. A., Mantgem, P. J. van, Middleton, B. R., Safford, H. D., Schwartz, M. W., & Trouet, V. (2023). Drivers of California's changing wildfires: A state-of-the-knowledge synthesis. *International Journal of Wildland Fire*, *32*(7), 1039–1058. https://doi.org/10.1071/WF22155

McClure, C. D., & Jaffe, D. A. (2018). US particulate matter air quality improves except in wildfire-prone areas. *Proceedings of the National Academy of Sciences*, *115*(31), 7901–7906. https://doi.org/10.1073/pnas.1804353115

Mirabelli-Montan, Y. A., Marangon, M., Graça, A., Mayr Marangon, C. M., & Wilkinson, K. L. (2021). Techniques for Mitigating the Effects of Smoke Taint While Maintaining Quality in Wine Production: A Review. *Molecules*, *26*(6), Article 6. https://doi.org/10.3390/molecules26061672

Mote, P. W., Li, S., Lettenmaier, D. P., Xiao, M., & Engel, R. (2018). Dramatic declines in snowpack in the western US. *Climate and Atmospheric Science*, *1*(1), Article 1. https://doi.org/10.1038/s41612-018-0012-1

National Centers for Environmental Information. (2020). *National Centers for Environmental Information (NCEI) 2020 National Temperatre and Precipitation Maps* [Dataset]. https://www.ncei.noaa.gov/access/monitoring/us-maps/?maps%5B0%5D=statewide-tavg-rank--3--202008

NOAA. (2011, November). *Monthly Climate Reports | National Climate Report | October 2011 | National Centers for Environmental Information (NCEI)*. https://www.ncei.noaa.gov/access/monitoring/monthly-report/national/201110#WRCC

Null, J., & Mogil, H. M. (2010). The Weather and Climate of California. *Weatherwise*, *63*(2), 16–23. https://doi.org/10.1080/00431671003603773

O'Dell, K., Ford, B., Fischer, E. V., & Pierce, J. R. (2019). Contribution of Wildland-Fire Smoke to US PM2.5 and Its Influence on Recent Trends. *Environmental Science & Technology*, *53*(4), 1797–1804. https://doi.org/10.1021/acs.est.8b05430

Pathak, T. B., Maskey, M. L., Dahlberg, J. A., Kearns, F., Bali, K. M., & Zaccaria, D. (2018). Climate Change Trends and Impacts on California Agriculture: A Detailed Review. *Agronomy*, *8*(3), Article 3. https://doi.org/10.3390/agronomy8030025

PRISM Climate Group. (2024). *PRISM Climate Group* [Dataset]. Oregon State University. https://prism.oregonstate.edu/

Santos, J. A., Malheiro, A. C., Karremann, M. K., & Pinto, J. G. (2011). Statistical modelling of grapevine yield in the Port Wine region under present and future climate conditions. *International Journal of Biometeorology*, *55*(2), 119–131. https://doi.org/10.1007/s00484-010-0318-0

Sheppard, S. I., Dhesi, M. K., & Eggers, N. J. (2009). Effect of Pre- and Postveraison Smoke Exposure on Guaiacol and 4-Methylguaiacol Concentration in Mature Grapes. *American*

*Journal of Enology and Viticulture*, *60*(1), 98–103. https://doi.org/10.5344/ajev.2009.60.1.98

Shertz, R. D., Kender, W. J., & Musselman, R. C. (1980). Effects of ozone and sulfur dioxide on grapevines. *Scientia Horticulturae*, *13*(1), 37–45. https://doi.org/10.1016/0304-4238(80)90020-5

Summerson, V., Gonzalez Viejo, C., Pang, A., Torrico, D. D., & Fuentes, S. (2021). Review of the Effects of Grapevine Smoke Exposure and Technologies to Assess Smoke Contamination and Taint in Grapes and Wine. *Beverages*, *7*(1), 7. https://doi.org/10.3390/beverages7010007

Tai, A. (2020). *PM2.5 & PM10 2025 Sequential Sampler Standard Operating Procedure*. Department of Ecology State of Washington.

United States Environmental Protection Agency. (2025). *Air Quality System (AQS)* [Data and Tools]. https://www.epa.gov/aqs

US EPA, O. (2019, August 13). *Why Wildfire Smoke is a Health Concern* [Overviews and Factsheets]. https://www.epa.gov/wildfire-smoke-course/why-wildfire-smoke-health-concern

Westerling, A. L., & Bryant, B. P. (2008). Climate change and wildfire in California. *Climatic Change*, *87*(S1), 231–249. https://doi.org/10.1007/s10584-007-9363-z

Whitnall, S., & Alston, J. (2025). Climate, weather, and collective reputation: Implications for California's wine prices and quality. *Journal of Wine Economics*, *20*(2), 122–167. https://doi.org/10.1017/jwe.2025.8

Wine Institute. (2023). *CALIFORNIA VINTNERS 2023 HARVEST REPORT*.

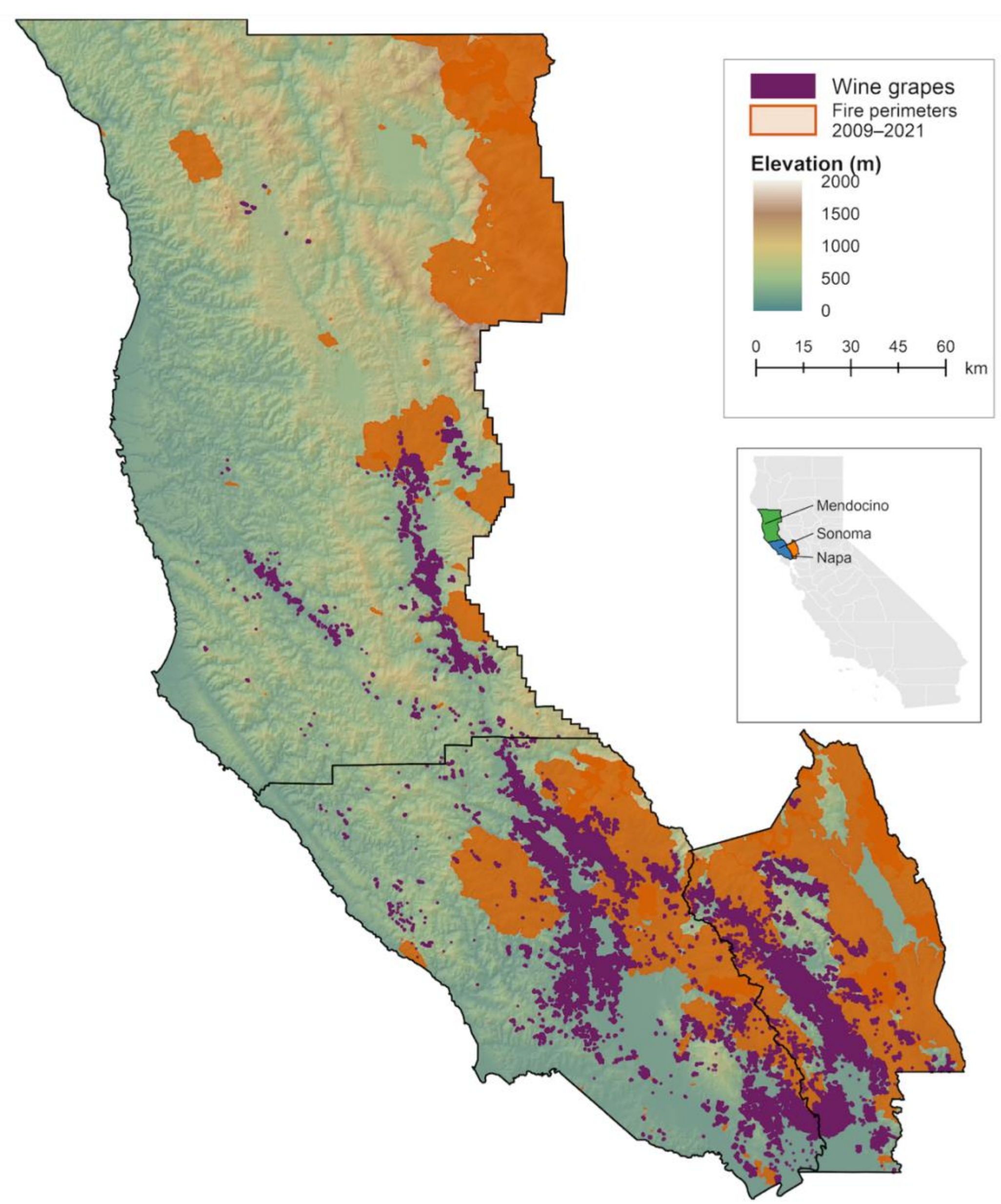

**Figure 1.** Study region encompassing Sonoma, Napa, and Mendocino counties (within California's Northen Coast wine producing region). Distribution of wine grape cropland across all three counties depicted in purple. Shaded fire perimeters (orange) indicate areas burned between 2009-2021.

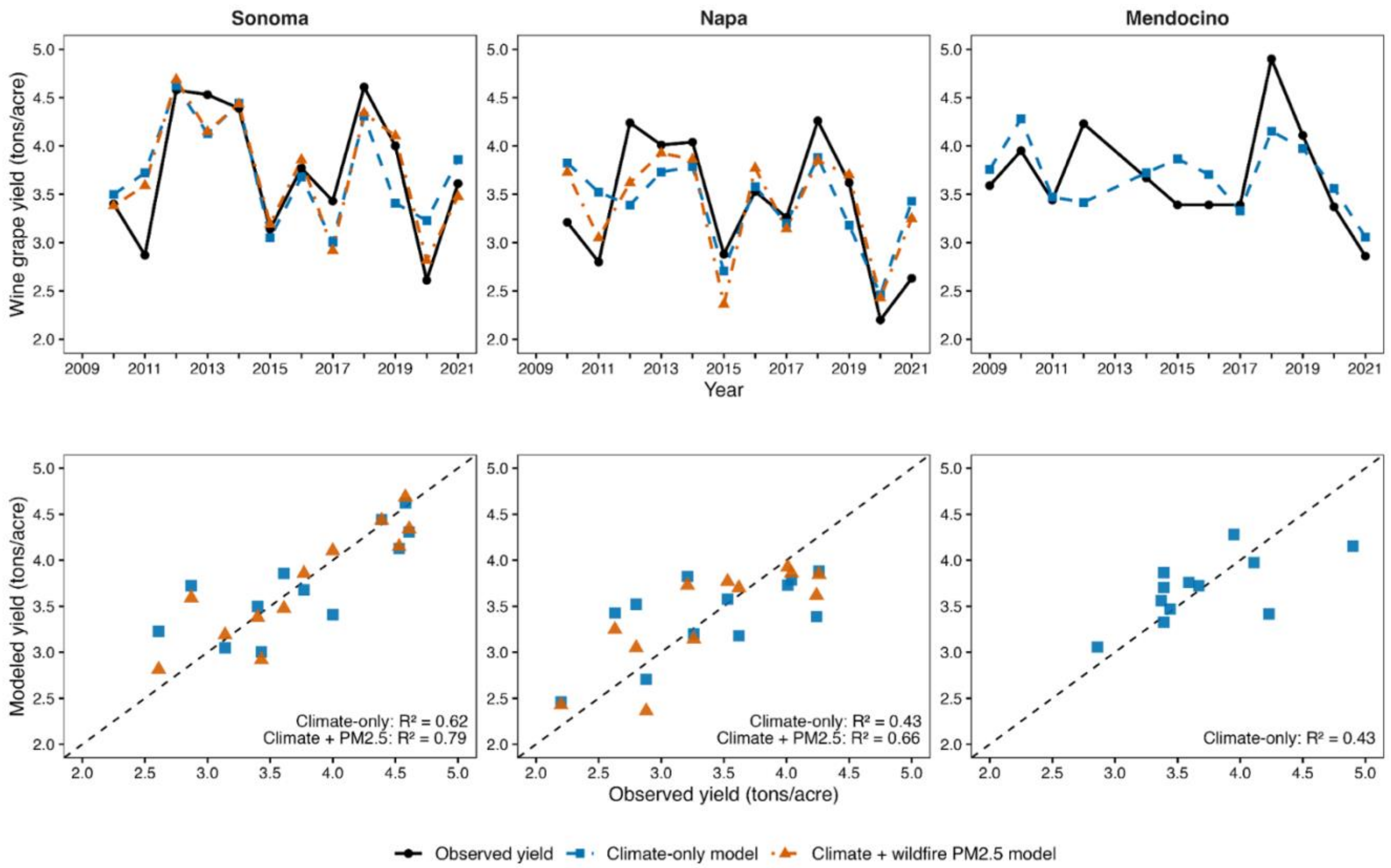

**Figure 2.** Observed and modeled wine grape yields for Sonoma, Napa, and Mendocino counties for duration of study period. Top panels show observed wine grape yields (black) compared to modeled yields for the climate variables only model (blue) and the climate and $PM_{2.5}$ variables model (orange). Bottom panels show the relationship between modeled and observed yields for each model specification. Concentration data was not available for Mendocino and was therefore omitted.

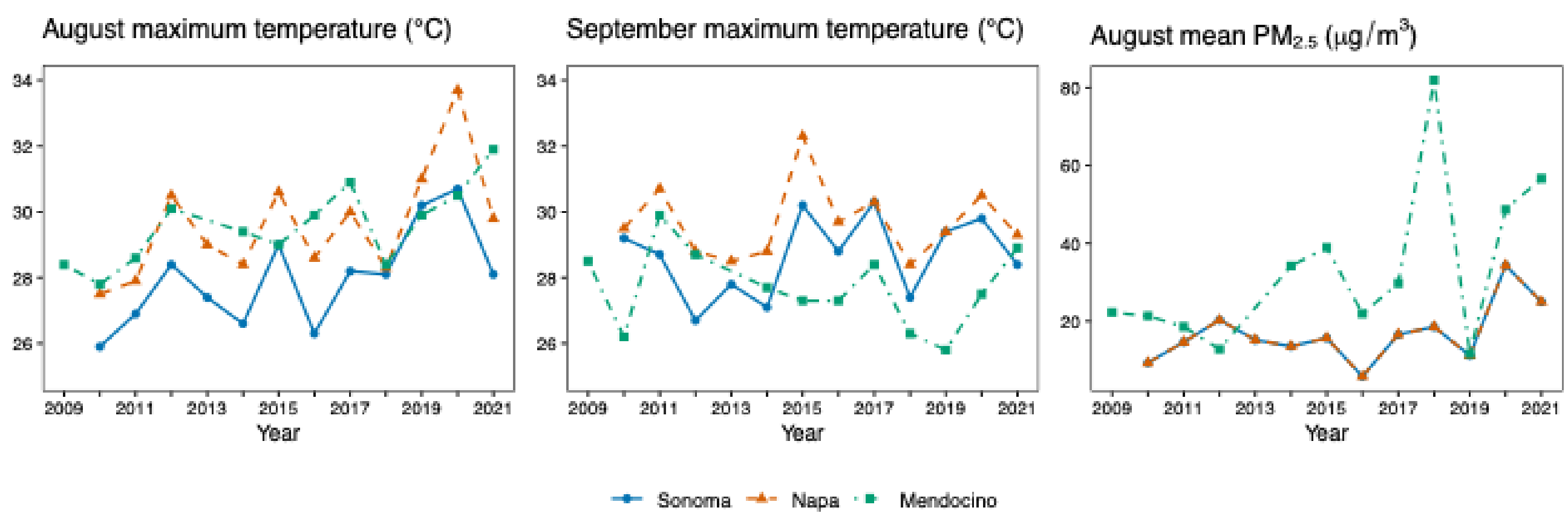

**Figure 3**. Sonoma, Napa, and Mendocino predictor variables as a function of years. Maximum monthly temperatures for Sonoma, Napa, and Mendocino in August (right panel) and September (center panel). Average of the monthly maximum $PM_{2.5}$ concentrations for Sonoma, Napa, and Mendocino in the month of August (left panel). Concentration data was not available for Mendocino and was therefore omitted.

## Tables:

**Table 1**. P-values for relevant predictor variables for Sonoma, Napa, and Mendocino counties. $T_{(harv)}$ denotes the monthly maximum (largest daily high) August and September temperatures divided by two at each site. Model results show that monthly average $PM_{2.5}$ (μg/m$^3$), maximum temperature and harvest temp have p-values below 0.10.

| County | Predictor Variable | P-Value |
|---|---|---|
| Sonoma | $T_{(aug)}$ | 0.0031 |
| | $T_{(sept)}$ | 0.0024 |
| | $T_{(harv)}$ | 0.0008 |
| | $PM_{(aug)}$ | 0.0350 |
| Napa | $T_{(sept)}$ | 0.0072 |
| | $T_{(harv)}$ | 0.0200 |
| | $PM_{(aug)}$ | 0.0731 |
| Mendocino | $T_{(harv)}$ | 0.0203 |

**Table 2.** Coefficient of determination ($R^2$) and adjusted $R^2$ (Adj. $R^2$) for regression models that predict yield from concentration and climatic data. Where Y represents modeled yield; monthly average $PM_{2.5}$ (PM), and monthly maximum temperature in Celsius (T). $T_{(harv)}$ denotes the monthly maximum (largest daily high) August and September temperatures divided by two at each site.

| County | Equation | $R^2$ | Adj. $R^2$ |
| --- | --- | --- | --- |
| **Sonoma** | $Y = 13.84 - 1.06\,T_{(harv)} + 0.75\,T_{(aug)} - 0.05\,PM_{(aug)}$ | 0.79 | 0.71 |
| | $Y = 16.63 - 0.45\,T_{(sept)}$ | 0.62 | 0.58 |
| **Napa** | $Y = 14.63 - 0.38\,T_{(harv)}$ | 0.43 | 0.38 |
| | $Y = 16.06 - 0.41\,T_{(sept)} - 0.03\,PM_{(aug)}$ | 0.66 | 0.58 |
| **Mendocino** | $Y = 13.98 - 0.34\,T_{(harv)}$ | 0.43 | 0.38 |